\documentclass[fleqn,usenatbib,onecolumn]{mnras}
\usepackage[utf8]{inputenc}
\usepackage{epsfig,amsmath,amssymb,subfigure}
\usepackage{blindtext}
\usepackage[ugly]{nicefrac}
\usepackage{hhline}
\usepackage[table]{xcolor}

\def\Aa{{\it Astron. Astrophys.} \,}

\def\apj{{\it ApJ \,}}

\def\php{{\it Phys. Plasmas}\,}

\def\apjl{{\it Ap. J. Lett.} \,}

\def\prd{{\it Phys. Rev. D.} \,}
\def\prl{{\it Phys. Rev. L.} \,}
\def\mn{{\it MNRAS} \,}

\def\rmph{{\it Rev. Mod. Phys.} \,}

\def\gr{$\gamma$-ray\,}
\def\xr{X-ray\,}

\def\ea{\ et al. \,}
\def\eg{{\it e.g. \,}}
\def\be{\begin{equation}}
\def\ee{\end{equation}}

\def\g{\gamma}

\def\pe{proton and electron }

\DeclareMathOperator{\erf}{erf}

\title[Energetic Particles in NGC253]{Energetic Particles in the Central Starburst, Disc, and Halo of NGC253}

\author[Yoel Rephaeli, Sharon Sadeh]{Yoel Rephaeli$^{1,2}$\thanks{E-mail: yoelr@tauex.tau.ac.il}, Sharon Sadeh$^{1}$\\
$^{1}$School of Physics and Astronomy, Tel Aviv University, Tel Aviv, 69978, Israel\\
$^{2}$Center for Astrophysics and Space Sciences, University of California,
San Diego, La Jolla, CA 92093-0424, USA}

\date{Accepted XXX. Received YYY; in original form ZZZ}

\begin{document}

\maketitle

\label{firstpage}

\begin{abstract}

Detailed modelling of the spectro-spatial distributions of energetic electrons and protons
in galactic discs and haloes of starburst galaxies (SBGs) is needed in order to follow
their interactions with the magnetized interstellar medium and radiation fields, determine 
their radiative yields, and for estimating their residual spectral densities in intergalactic environments. We have developed a semi-analytical approach for calculating the particle spectro-spatial distributions in the disc and halo based on a diffusion model for particle propagation from acceleration sites in the central SB and disc regions, including all their relevant interaction modes. Important overall normalization of our models is based on previous modelling of the Galactic disc (with the GALPROP code), scaled to the higher star-formations rate in NGC253, and on spatially resolved radio measurements of the central SB and disc.
These provide the essential input for determining the particle distributions and their predicted radiative yields in the outer disc and inner halo for a range of values of the key parameters that affect diffusion rate and energy losses. Results of our work clearly indicate that quantitative description of non-thermal emission in SBGs has to be based on modelling of the particle distributions in the entire disc, not just the central SB region.

\end{abstract}


\section{Introduction}

Emission by energetic particles (`cosmic rays') in some of the nearby star-forming galaxies 
has been measured extensively over a sufficiently wide spectral range that provides
a reasonably good basis for detailed studies of the particle spectro-spatial distributions
(SSDs) in interstellar space. Such studies are particularly well motivated in the case of 
several galaxies from which non-thermal (NT) emission has been measured not only in 
radio and X-ray, but also at $\gamma$-ray energies, mostly by the Fermi satellite (e.g., 
Andromeda, M82, and NGC253; Abdo \ea 2010a; Abdo \ea 2010b). The latter two galaxies are of
special interest due to their relatively high star-formation rate (SFR) that directly
translates to more intense NT emission by a higher flux of energetic particles than in 
`normal' spirals. Indeed, these galaxies are the closest examples of starburst galaxies 
(SBGs), in which the enhanced SFR and SN activity are manifested by intense optical and IR emission.

Interest in studies of NT emission in SBGs began long ago with quantitative modelling
the radio and hard X-ray emission by energetic electrons in the disc and halo of NGC253
(e.g., Goldshmidt \& Rephaeli 1995); this was expanded to include high energy emission 
by electrons and protons (Paglione \ea. 1996; V\"{o}lk, Aharonian \& Breitschwerdt 1996; Torres 2004;
Domingo-Santamar\'{i}a \& Torres 2005), followed by detailed predictions of the high 
energy $\gamma$-ray spectra of both M82 and NGC253 by Persic, Rephaeli, \& Arieli 
2008, de Cea del Pozo et al. 2009, Rephaeli, Arieli, \& Persic 2010, Lacki \ea 2011.
Indeed, the levels of \gr emission predicted in some of these works were proven to be 
quite realistic when these galaxies were detected in the GeV--TeV range by Fermi,
VERITAS, and H.E.S.S telescopes (Abdo \ea 2010a; Acciari et al. 2009; Acero et al. 2009).

Main objectives in most of the studies of energetic electrons and protons in SBGs were 
the spectra and levels of (low and high) energy \gr emission in the galactic disc.
Given that in SBGs a significant fraction of the total SF activity takes place in a relatively 
small and dense central (`nuclear') region, in some previous calculations quantitative 
predictions of NT emission were based on the assumption that \gr emission is dominated by 
energetic protons in the central SB (CSB) region. Due to the relative smallness of this 
compact region, all relevant quantities characterizing the particles and media -- gas density, magnetic and radiation fields -- were assumed to be uniform. The main objective of such an approximate treatment was estimation of the high energy \gr emission, based on the presumption that the high SFR and gas density in the CSB region imply that most of the total galactic GeV--TeV emission is produced by neutral ($\pi^{0}$) decays as result of energetic proton interactions with ambient gas protons in this central region. Clearly, for a more
realistic and useful description of the particles and their radiative yields, a detailed 
spectra-spatial description is required in order to account for the different distributions of particle acceleration sites and effective propagation mode in the central CSB and disc
regions, as well as an assessment of the relative strengths of the radio, hard \xr, and \gr emission from these regions. 

This approximation was shown to be quantitatively inadequate when emission from the 
full galactic disc was modelled based on the GALPROP (e.g., Moskalenko \ea 2003) code.
A  modified version of this code was developed to predict the spectral energy distribution of M82 (Persic \ea 2008) and NGC 253 (Rephaeli, Arieli, \& Persic 2010). Normalizing the particle spectra by the SFR and the measured radio spectra in the central and full disc regions, it was found that a significant fraction of the total galactic \gr emission in M82 and NGC 253 comes from outside the CSB region. It is thus clear that a more realistic description is required of the particle and (magnetic and radiation) fields distributions across the full galactic disc; whereas an approximate model could be sufficient in order to estimate the overall level of \gr emission, mostly for assessing the likelihood of its detection (originally, by \textit{Fermi}/LAT), detailed analyses of radio and \gr measurements necessitate modelling of the full particle SSD in the disc.

The intense SF and SN activity, the efficient acceleration of electrons and protons, and 
their coupling to IS media, radiation and magnetic fields, motivated a more comprehensive evolutionary approach to the study of NT phenomena in SBGs, one that is based on magnetohydrodynamical (MHD) simulations. This necessarily more detailed approach
that, in principle, enables a more complete description of relevant hydrodynamical, 
thermal, and NT processes, is clearly more ambitious and challenging than the above-mentioned empirical (largely numerical) approach that focuses only on energetic particles
and their interactions with magnetic and radiation fields and coupling to (non-evolving) 
IS media. Simulation-based studies of energetic particles and their radiative yields were 
carried out, among others, by Chan \ea (2019), Hopkins \ea (2020), and Werhahn \ea, 
(2021, 2023). 

MHD simulations could possibly afford a more realistic description of the complexity 
of physical phenomena in IS media, and -- in particular -- the various plasma processes that couple energetic electrons and protons to the magnetized gas and determine their modes of propagation (e.g., Ruszkowski \ea 2017, Thomas \ea  2023). While these studies can provide
a more detailed, relatively fine-grained description of the particle SSDs as compared with the empirical (necessarily coarse-grained) approach, inherent uncertainties
in the description of the evolutionary history of the galaxy on the one hand, and the limited quality of observational spectro-spatial data (even from nearby galaxies), imply that both approaches are not only viable, but also can provide useful (and possibly complementary) insight. 

We continued our empirically based study of energetic electron and proton SSDs and their radiative yields in the disc and halo of a SFG by developing a new semi-analytical approach based on a diffusion model for particle propagation from acceleration sites and interactions
in both disc and halo. Important overall normalization of the particle SSDs was set by the observationally estimated SFR, similarly to what was done in global modelling of the Galaxy with the GALPROP code (Strong \ea 2010). Fitting predicted radio spectra to the measured spectral and spatial profiles of radio measurements fully determined the SSDs in the outer disc and provided a reasonably reliable basis for predicting the particle distributions and radiative yields in the halo for a range of values of key parameters affecting energy losses and diffusion. In the first analysis using this semi-analytical approach we (Rephaeli and Sadeh 2019) selected NGC 4631 and NGC 4666, two of 35 nearby edge-on SFG included in the EVLA radio continuum survey CHANG-ES (Wiegert \ea 2015), which yielded the first co-added median map of the halo of a spiral galaxy at 1.5 GHz, with emission extending to more than 20 kpc above the disc. Doing so enabled not only the determination of key NT quantities across the full disc, but also provided a more adequate basis for extending the quantitative description of the particle SSD to the galactic halo.

For a realistic description of energetic particles in a spiral galaxy with a high SFR, 
and in which a significant fraction of the SF activity occurs in a compact CSB region, 
improved modelling of the particle SSDs and radiative yields is needed to account for
the much higher particle fluxes, magnetic and radiation fields. The elevated values of 
all these quantities in the CSB are reflected in a much more intense emission in essentially 
all spectral bands than would otherwise be predicted from the central part of a one-zone disc.
This need is even stronger in the case of the two nearby well probed SBGs M82 
and NGC 253. Clearly, the nearly optimal capability of spectro-spatial radio mapping of 
these galaxies, and in light of future spatially resolved measurements in high X-and-\gr measurements, it is quite essential to model NT emission in the CSB, disc, and halo regions of these iconic SBGs.

Here we apply our modified multizone code to the study of NT emission in NGC 253.
In contrast to M82, NT emission in NGC 253 is thought to be largely a result of 
stellar-driven activity, with only a minor contribution by a massive blackhole (MBH)
in the galactic centre. As such, it is a better suited example for the study of NT
phenomena in a SBG. We describe \pe SSDs in the CSB, disc, and halo regions of
NGC 253 based on our diffusion-based model for their propagation as they traverse 
the magnetized gas and lose energy by all relevant processes. In Section 2, we present
the basic model and results of its implementation. Our quantitative treatment provides 
a reliable basis for predicting the levels of radio, X-\&-$\gamma$-ray emission in all 
three regions, as detailed in Section 3, and further discussed in Section 4.

\section{Methodology}

It is well established that particle acceleration in SFGs is largely a stellar-related phenomenon 
driven by SN shocks and pulsar wind nebulae. Traversing the multiphase IS media the particles lose energy by interactions with the magnetized IS gas and radiation fields,
eventually reaching the dilute halo where their propagation is much faster and energy losses much weaker than in the disc. As stated above, we treat separately the CSB and disc-halo regions. Full treatment of the radio and hard X-ray emission produced by the interaction of primary and secondary electrons with the ambient magnetic fields, and of the $\gamma$-emission produced by the decay of $\pi^0$ mesons created in hadronic collisions between energetic protons and galactic gas, is presented in full details elsewhere (Rephaeli and Sadeh, 2019).

We focus on the main NT properties of the nearby SBG NGC 253, namely the particle 
source distribution, the magnitude and morphology of the magnetic field, gas density, 
and diffusion coefficient in particular, as well as the structure of the visual and far-infrared (FIR) radiation fields in the CSB and disc. It should be noted that in the
common diffusion-based approach of energetic particle propagation in IS space, the 
diffusion coefficient is an effective quantity in the description of what realistically is a 
more complex propagation process that may also include other modes of transport,
such as convection and streaming (e.g., Zweibel 2013), possibly as part of a galactic wind. 
Our approach here is necessarily simplified due to the complexity of IS media and the lack 
of higher sensitivity and spatial resolution measurements. Accordingly, the effective 
diffusion process along the direction perpendicular to the disc is taken to be constant in the small CSB region, spatially variable in the disc and halo. In addition to spatial dependence, diffusion could also depend appreciably on the particle energy. Therefore, we have expanded our previous treatment -- which was based on energy-independent diffusion -- to include also
the case when the diffusion coefficient varies with energy in the form deduced by Werhahn 
\ea (2021) based on simulations with the AREPO MHD code of a sample of model galaxies. The predicted \gr spectra from energetic protons and electrons in these simulations were normalized to \gr observations of three nearby galaxies that included NGC 253 (in addition
to M82 and NGC 2146).

\subsection{Central SB region}

We model the CSB region as a sphere with radius 300 pc (Kapi\'{n}ska \ea 2017) and uniform 
magnetic field, gas density, and diffusion coefficient. Particle sources are assumed to follow a 
Gaussian distribution with variance $\sigma_R=$100 pc. Specifically,
\begin{eqnarray}
f(r_i)=\frac{4\pi}{(2\pi)^{3/2}}\frac{1}{\sigma_R^3}\times
e^{-r_i/2\sigma_R^2}, 
\end{eqnarray}
where $r_i$ is the radial position of source $i$. Integrating the Green function solution of the 
diffusion equation (e.g.\,Atoyan, Aharonian, \& V\"{o}lker 1995) over all sources we obtain 
the spatio-temporal spectral distribution of primary electrons and protons:
\begin{eqnarray}
 f_{e,p}(t,\gamma,r_o)&=&
\frac{\Delta N(\gamma_t)P(\gamma_t)}{\pi^{3/2}P(\gamma)r_d^{3}}
\frac{1}{\sigma_R^3}\times e^{-r_o^2/r_d^2} \\ \nonumber
&\times& e^{2r_o^2 r_s^2/(r_d^4+2r_d^2\sigma_R^2)}\left(2/r_d^2+1/\sigma_R^2\right)^{-3/2},
\end{eqnarray}
where $r_s$, $r_o$, and $r_d$ are, respectively, the CSB radius, observation point, and diffusion radius; $\gamma_t$ and $\gamma$ are the pre- and post-cooling energies of the diffusing particle, $\Delta N$ is the particle spectral injection rate, and P($\gamma$) is the energy loss-rate. The latter expression is convolved with the synchrotron and Compton (Blumenthal \& Gould 1970), and  $\pi^0$ decay (Mannheim \& Schlickeiser 1994) emissivities, then time-integrated to obtain the full spatio-temporal spectral radiation emissivity at distance $r_o$, observation time $t_o$, and radio frequency $\nu$, or
$\gamma$-photon energy E$_{\gamma}$. To calculate the radio, hard X-ray, and 
$\gamma$ radiative yields the magnetic field, $B$, gas density, $n$, and diffusion 
coefficient, $D$, need to be selected. We consider three cases differing by their diffusion coefficient and the power index of the particle injection spectrum, $\alpha$, as specified in table~(\ref{tab:tabpar_nuc}). Following Strong \ea (2010), the fiducial particle injection spectrum is assumed to be a power-law with index $\alpha=1.8$ below break energy of
4 GeV, and 9 GeV for electrons and protons, respectively, and $\alpha=2.25$ above the
break.

Whereas the diffusion coefficient is assumed to be spatially uniform in this (small) region, 
we consider both cases of energy-dependent and energy-independent diffusion (hereafter, 
EDD and EID). Given that EDD is physically motivated (e.g., Evoli
\ea 2020), the EID case can be regarded as a limiting case of very weak energy dependence. We adopt the scaling $D=D_0(E/E_0)^{\delta}$, with E$_0$=3 GeV, and $\delta$=0.3 (Werhahn \ea 2021); the latter value was found to provide a better fit to results of the authors' (AREPO) simulations than the somewhat higher value ($0.5$) that is assumed sometimes (e.g., Evoli \ea 2020). We note that in comparison with EID, diffusion is
slower (faster) below (above) E$_0$, namely $\gamma_e\sim$5900 and $\gamma_p \sim$4 for electrons and protons, respectively. This will be useful in elucidating the differences between the calculated radiative yields in the EDD and EID models.

The optical and FIR radiation fields are taken to be uniform greybodies at temperatures
5000$^{\circ}\,$K and 30$^{\circ}\,$K, with emissivities $\sim$2.6 $\times 10^{-14}$ and
0.017, respectively.

\begin{center}
\begin{table}
\begin{tabular}{|l|c|c|c|} \hline
& C1 & C2 & C3 \\ \hline
n (cm$^{-3}$) & 110 & 110 & 110  \\ \hline
B ($\mu$G) & 160 & 160 & 160 \\ \hline
D (cm$^2$ s$^{-1}$) & $1\times 10^{28}$ &
$1\times 10^{28}$ & $3\times 10^{28}$ \\ \hline
$\alpha$ & 1.8 & 1.5 & 1.6 \\ \hline 
\end{tabular}
\caption{\label{tab:tabpar_nuc} 
CSB parameters sets, cases C1--C3: gas density (n), magnetic field (B), diffusion coefficient (D) -- all taken to be spatially uniform across the CSB region, and power index of the particle injection spectrum.}
\end{table}
\end{center}

\subsection{Disc-halo region}

It is estimated that $\sim$ 70\% of the energetic particle sources in NGC 253 are in the CSB region where the estimated SFR is $\sim 5$ M$_{\odot}$yr$^{-1}$ (Wik\ea 2014). The total source distribution in the full disc is represented as a sum of two Gaussians with a relative normalization corresponding to this fraction of sources in the central region.

\begin{eqnarray}
 f(x_i,y_i,z_i)=&&\,\,\,\,k_1\frac{1}{(2\pi)^{3/2}}
 \frac{1}{\sigma_R^3}\exp{\left[-(x_i^2+y_i^2+z_i^2)/2\sigma_R^2\right]}\\ \nonumber
 &&+k_2\frac{1}{\pi^{3/2}}\frac{1}{\sigma_a^2\sigma_c}
 \exp{\left[-(x_i^2+y_i^2)/\sigma_a^2-z_i^2/\sigma_c^2\right]},
\end{eqnarray}
where $\sigma_a=12$ kpc (somewhat smaller than $\sim$13.8 kpc, the disc radius inferred
by Lucero \ea 2015) and $\sigma_c=1$ kpc is a measure of the scale height of the galactic
disc, and the constants k$_1$ and k$_2$ are determined by the relative fraction of sources
in the CSB,  
\begin{eqnarray}
&&\int_{-\sigma_R/2}^{\sigma_R/2}\int_{-\sigma_R/2}^{\sigma_R/2}\int_{-\sigma_R/2}^{\sigma_R/2}f(x_i,y_i,z_i)dx_idy_idz_i=0.7\\ \nonumber
&&\int_{-\infty}^{\infty}\int_{-\infty}^{\infty}\int_{-\infty}^{\infty}f(x_i,y_i,z_i)dx_idy_idz_i=1,
\end{eqnarray}
with the solution
\begin{eqnarray}
&k_1=&\frac{0.7-k_2\erf^2\left(\frac{\sigma_R}{2\sigma_a}\right)
\erf\left(\frac{\sigma_R}{2\sigma_c}\right)}
{\erf^3\left(\frac{1}{2\sqrt{2}}\right)} \\
&k_2=&\frac{\erf^3{\left(\frac{1}{2\sqrt{2}}\right)}-0.7}
{\erf^3\left(\frac{1}{2\sqrt{2}}\right)-\erf^2\left(\frac{\sigma_R}{2\sigma_a}\right)
{\erf\left(\frac{\sigma_R}{2\sigma_c}\right)}},
\end{eqnarray}
where $\erf$ is the error function.

\subsubsection{Key physical quantities}

To account for the spatial variation in the relevant quantities, we describe the
variability by (integrable) functions that approximate the complex disc morphology.
We model the spatial distributions of key quantities based on the observationally 
determined radio morphology of the disc and inner halo of (an edge on) spiral galaxy.
As noted in the Introduction, this was achieved by Wiegert \ea (2015) from co-added 
1.5 GHz maps of 30 nearby SFGs observed as part of the CHANG-ES EVLA survey. 
The median map of the emission profile well beyond the galactic disc provides a
quantitative basis for modelling the spatial distributions of the relevant quantities that
affect the propagation of energetic electrons and protons in the disc and halo -- gas density, magnetic field, and diffusion coefficient. The disc-halo region is described by a sequence
of concentric ellipsoidal shells, each with a uniform gas density, magnetic field, and 
diffusion coefficient. The ellipsoids are taken to be highly eccentric towards the centre, gradually becoming spherical at $\sim 20$ kpc, represented in terms of the function
\begin{eqnarray}
 e_i=\frac{e_1}{\exp{[(a_i-15)/.5]}+1},
\end{eqnarray}
where $e_1=\sqrt{1-(c/a)^2}\approx 0.96$, $a=12$ kpc, $c=3.4$ kpc (Heesen \ea 2009), and $a_i$ is a 500-element vector in the plane of the disc extending from the galactic centre
to a radial distance of 350 kpc. The vectors a$_i$ and e$_i$ fully determine the sequence of 500 concentric ellipsoids, the surfaces of which are expressed by $(x^2+y^2)/a_i^2+z^2/c_i^2=1$, where $c_i=a_i\sqrt{1-e_i^2}$.

Next, the gas density, magnetic field, and diffusion coefficient have to be specified in each 
shell. Since for each point on a given shell $B$, $n$, and $D$, are constants by construction, 
it is convenient to associate with every shell their values at the shell intersection with the 
$x$-axis, i.e. where ($x$,$y$,$z$)=($a_i$,\,0,\,0). We describe the spatial distribution of these quantities
by $\beta$-King-like profiles, with the physically motivated scaling B$\propto n^{2/3}$, 
and with 3 vectors of 500 elements specified each at the shell-$x$-axis intersection:
\begin{equation}
n(a_i)=n_0/[1+(a_i/r_c)^2]^{\beta},
\end{equation}
\begin{equation}
B(a_i)=B_0/[1+(a_i/r_c)^2]^{2\beta/3},
\end{equation}
and the diffusion coefficient, in both EDD and EID cases, is assumed to increase as
\begin{equation}
D(a_i)=D_0[1+(a_i/r_c)^2]^{\beta_D},
\end{equation}
where n$_0$, B$_0$, and D$_0$ are their central values and $r_c$, $\beta$, $\beta_D$ are 
treated as free shape parameters to be determined by comparison with measurements. These 
are supplemented by the disc and total gas mass, which is computed by integrating the gas
density profile over the respective volumes, using $n_0$, $r_c$ and $\beta$. 

We construct a 3D Cartesian grid with 50$\times$50$\times$100 elements in the ($x$,\,$y$,\,$z$) axes,
each of which extends out to 350 kpc. As with the CSB, the spatio-temporal distributions of primary electrons and protons is computed at each grid point as 
\begin{eqnarray}
f_{e,p}(t,\gamma,\overline{r}_o) =\int_{-\infty}^{\infty}
\int_{-\infty}^{\infty}\int_{-\infty}^{\infty}
\frac{\Delta N(\gamma_t)P(\gamma_t)}{\pi^{3/2}P(\gamma)r_d^{3}}
\left[\frac{k_1}{(2\pi)^{3/2}}\frac{1}{\sigma_R^2}
e^{-(x_i^2+y_i^2+z_i^2)/2\sigma_R^2}+
\frac{k_2}{\pi^{3/2}}\frac{1}{\sigma_a^2\sigma_c}
e^{-(x_i^2+y_i^2)/\sigma_a^2-z_i^2/\sigma_c^2}\right]dx_idy_idz_i.
\end{eqnarray}
These are convolved with the synchrotron, Compton, and $\gamma$ emissivities, and time-integrated to yield the full spatio-temporal spectral emissivity at distance $r_o$, observation 
time $t_o$, and radio frequency $\nu$, or $\gamma$-photon energy E$_{\gamma}$. 
($B$, $n$, and $D$, at each grid point are inferred by interpolating between their counterparts 
on values in the 2 adjacent ellipsoidal shells, as these do not necessarily coincide with the 
Cartesian grid.) For the computation of the radio- and $\gamma$-emissivities produced in the 
disc-halo region, we have considered 3 cases whose parameter sets are listed in table~(\ref{tab:tabpar_hal}). As with the CSB region, the optical and FIR radiation fields are
taken to be uniform greybodies at temperatures 5000$^{\circ}\,$K and 30$^{\circ}\,$K, with greybody emissivities $\sim$9.4$ \times 10^{-14}$ and 3$\times 10^{-5}$, respectively.
Outside the disc the radiation fields fall radially as r$^{-2}$.

\begin{center}
\begin{table}
\begin{tabular}{|l|c|c|c|} \hline
& D1 & D2 & D3 \\ \hline
n$_0$ (cm$^{-3}$) & 1.5 (1.5) & 1.5 (1.5)& 1.5 (1.5) \\ \hline
B$_0$ ($\mu$G) & 30 (20) & 30 (30) & 30 (50) \\ \hline
D$_0$ (cm$^2$ s$^{-1}$) & 3E28 (3E28)& 6E28 (3E28) &
3E28 (6E28) \\ \hline
$\beta$ & 1.3 (1.3)& 1.3 (1.3)& 1.3 (1.3) \\ \hline
$\beta_D$ & 0.4 (1.0)& 0.4 (1.1)& 0.5 (1.4) \\ \hline
M$_{g_d}$ ($\times 10^{9}$ M$_{\odot}$) & 2.8 (2.8)& 2.8 (2.8)& 5.2 (5.2) \\ \hline
M$_{g_t}$ ($\times 10^{9}$ M$_{\odot}$) & 21.7 (21.7)& 21.7 (21.7)& 45 (45) \\
\hline
\end{tabular}
\caption{\label{tab:tabpar_hal} 
Disc and halo parameter sets, cases D1--D3: central gas density (n$_0$), magnetic field (B$_0$), diffusion coefficient (D$_0$), and core radius (r$_c$). $\beta$ and $\beta_D$ of
the King $\beta$ type profiles are, respectively, the shape parameters of the gas density 
and diffusion coefficient. Also listed are the disc-halo (M$_{g_{d}}$), and the total
(M$_{g_{t}}$) gas masses, respectively. Figures in parentheses denote the respective parameters for EID coefficient.}
\end{table}
\end{center}

\section{Results}

Spectral radio, hard X-ray, and $\gamma$-emissivity grids were computed for both the CSB 
and disc-halo regions. The radio emissivities were calculated at 0.61, 1.41, 1.45, 1.47, 1.66,
4.52, 4.85, 4.89, 6.70, 7.00, and 8.09 GHz; the hard X-ray spectrum at 7--200 keV, and the $\gamma$-spectrum was calculated at each decade in the 1 MeV--10 TeV range. These
were integrated over the respective volumes and converted to fluxes taking 3.5 Mpc for the distance to NGC253, resulting in spectral radio, hard X-ray, and $\gamma$-fluxes. 

\begin{figure}
\centering
\epsfig{file=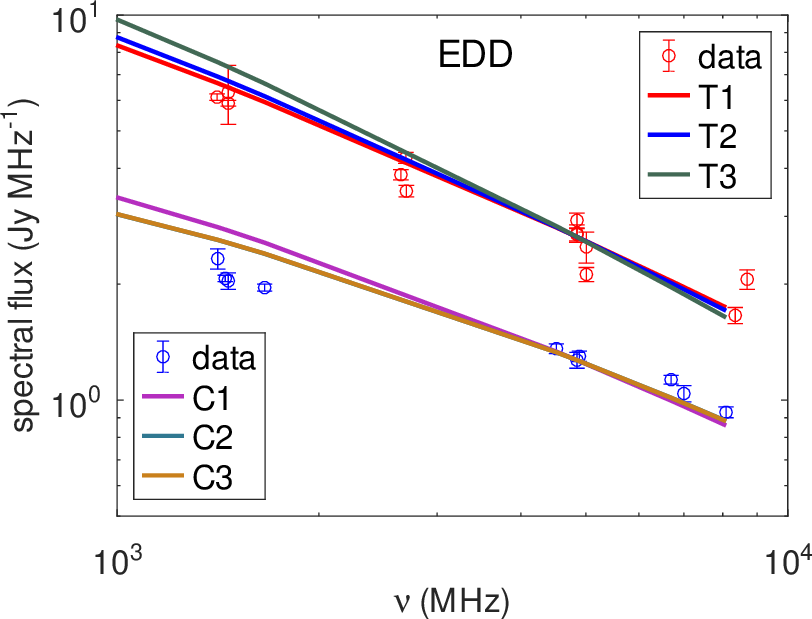,width=7.2cm,height=7.2cm,clip=}
\hskip -1mm
\epsfig{file=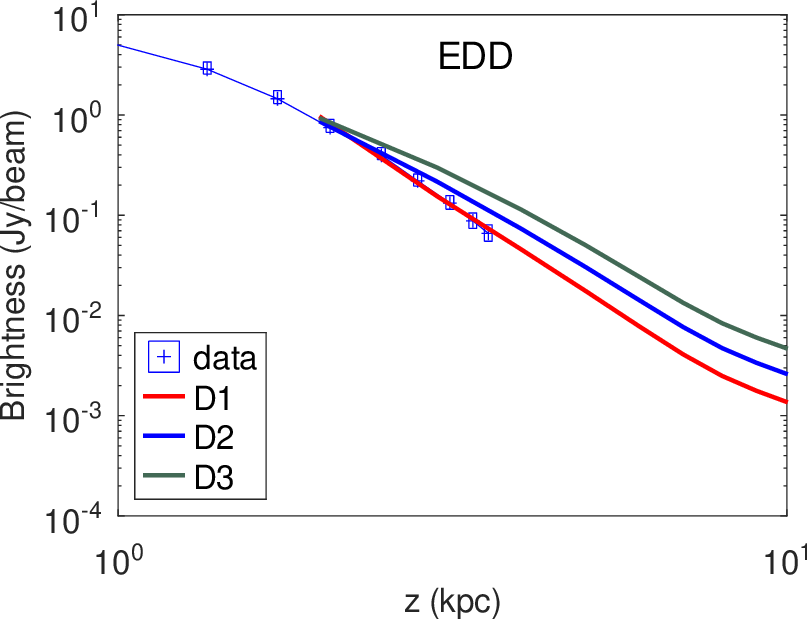,width=7.2cm,height=7.2cm,clip=}\\
\epsfig{file=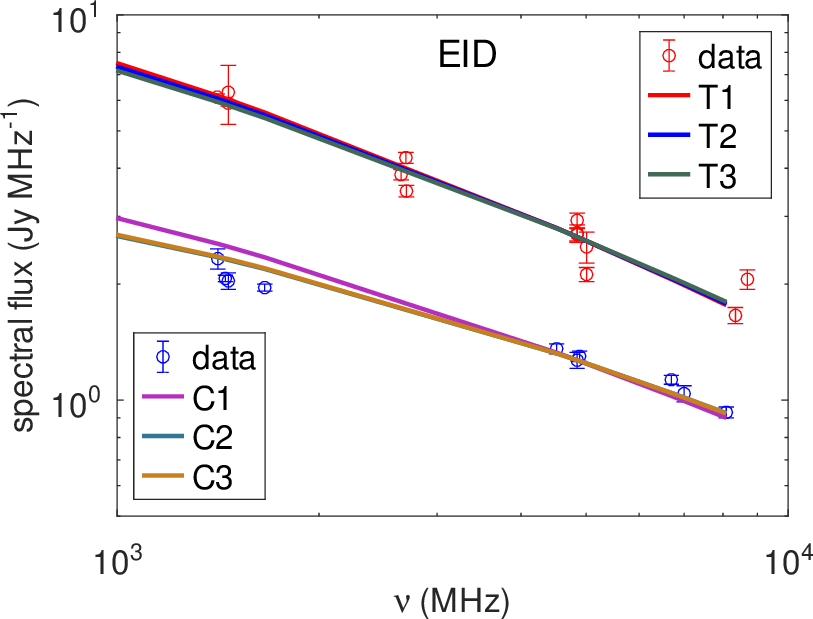,width=7.2cm,height=7.2cm,clip=}
\hskip -1mm
\epsfig{file=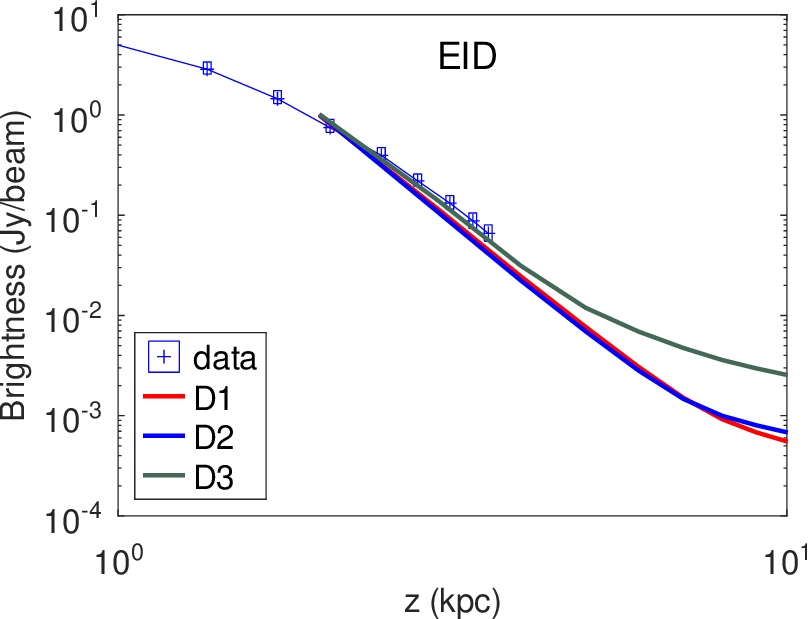,width=7.2cm,height=7.2cm,clip=}
\caption{Upper and lower left-hand panels: Radio spectra of the CSB (cases C1--C3) and disc-halo including the CSB (cases T1--T3), in spectral flux units, for energy-dependent and independent diffusion. Upper and lower right-hand panels: the measured (Kapi\'{n}ska \ea 2017) and predicted 227 MHz brightness profiles at altitudes $z\,\ge\,$.93 kpc above the galactic plane (cases D1--D3), for energy-dependent and independent diffusion.}
\label{fig:radio}
\end{figure}

Radio emissivity grid of the disc-halo was computed at 227 MHz and integrated along
the los at increasing heights above the galactic plane, then convolved with a Gaussian beam of
$\sigma_{B}$=1.7 arcmin, to produce spectral brightness profiles. Radio fluxes originating in 
the CSB and the disc-halo are plotted in the left-hand panels of Fig.~\ref{fig:radio}, together
with measured data compiled by Kapi\'{n}ska \ea (2017). Our results are normalized to the
level of the measured flux at 4.55 GHz, for comparison of the computed and measured spectra. 
The total computed to measured normalization factors for EDD vary from $\sim$ 0.5 to 2.1 (cases 1 and 3, respectively), whereas those calculated for EID are between $\sim$0.7
(case 1) and 2 (case 3). The corresponding normalization factors  for the CSB alone are
$\sim$1.4--2.6 and 0.5--1 for EDD and EID, respectively.

The calculated radio flux spectra in the EDD case (upper left panel) are marginally consistent with the measurements, while clearly steeper than in the EID case (lower left panel) which provides a better overall fit to the data. As noted in Section 2, the scaling of the diffusion coefficient with energy suggests that particles with energies lower than E$_0$=3 GeV, namely 
$\gamma\sim$5900 for electrons, diffuse slower than when D=D$_0$. With a central field value of $\sim$20--30$\mu$G, the synchrotron spectrum peaks around $\nu\sim$3--4 GHz; consequently, the lower frequency part of the spectrum originates mainly from lower energy electrons that spend longer times in the higher magnetic field (and denser) regions due to slower diffusion. While these electrons enhance the radio emission at lower frequencies, higher energy electrons diffuse more rapidly out of these regions and therefore contribute less to the higher frequency part of the spectrum. These considerations can explain the somewhat steeper EDD spectra. The right-hand panels of the figure describe the radio profiles of the
disc-halo for cases D1--D3, together with the measured profile inferred from Fig.~2 of Kapi\'{n}ska \ea (2017). Here too the calculated brightness profiles are normalized to the measured profile. Examination of the figure indicates that disc-halo case D1 best matches the measurements up to a height of $z\,\sim\,$3.6\,kpc above the disc midplane for EDD, whereas case D3 best fits the data for EID. Generally, and up to a certain altitude above the disc plane,
EDD profiles are somewhat flatter than the EID profiles since at 227 MHz most of the energetically relevant electrons diffuse slower than predicted in the EID case.

\begin{figure}
\centering
\epsfig{file=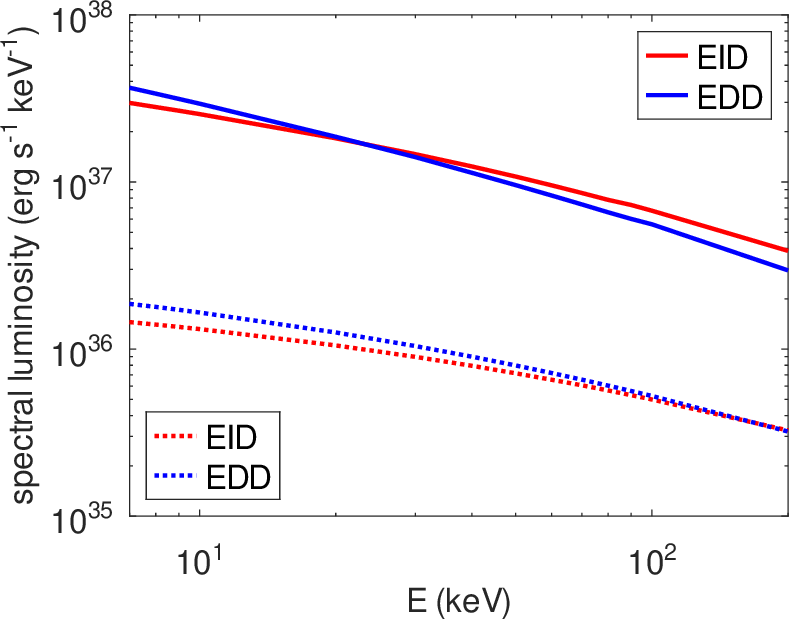,width=7.2cm,height=7.2cm,clip=}
\caption{Hard X-ray spectral flux calculated in the 7--200 keV range.
Shown are the CSB (dotted) and total (solid) spectra for case 1, with
energy-dependent and independent diffusion.}
\label{fig:hard_x}
\end{figure}

Spectral 7--200 keV luminosities of the CSB and the (entire) galaxy are shown in Fig.~\ref{fig:hard_x}. The integrated CSB 7--20 keV fluxes predicted in case C1 are
$\sim$1.1$\times$10$^{-14}$ erg s$^{-1}$cm$^{-2}$ for the EID model, and 1.4$\times$10$^{-14}$ erg s$^{-1}$cm$^{-2}$ for the EDD, both well below the 90$\%$ confidence upper limit of $\sim$1.8$\times$10$^{-13}$ erg s$^{-1}$cm$^{-2}$ deduced by Wik \ea (2014) from NuStar observations. This limit was obtained from an estimate of the measured residual emission after accounting for detected point sources in a central region roughly comparable in size to that of the CSB. The corresponding integrated CSB fluxes for cases C2 and C3 are also well below this limit. As shown in Fig.~\ref{fig:hard_x}, most of
the predicted hard X-ray emission originates outside the CSB; the 7--20 keV fluxes from the full disc are $\sim$2.1$\times$10$^{-13}$ erg s$^{-1}$cm$^{-2}$ (EID), and 2.3$\times$10$^{-13}$ erg s$^{-1}$cm$^{-2}$ (EDD). All the quantitative limits set by
Wik \ea (2014) are for varying central regions of up to $\sim 1$ kpc in radius. Our 
results imply that the levels of Compton-produced 7--20 keV emission in both the CSB
and the full disc are quite comparable in the EID and EDD models, with slightly steeper spectra in the latter case as a result of the somewhat steeper electron spectrum (due to $D \propto E^{0.3}$).

\begin{figure}
\centering
\epsfig{file=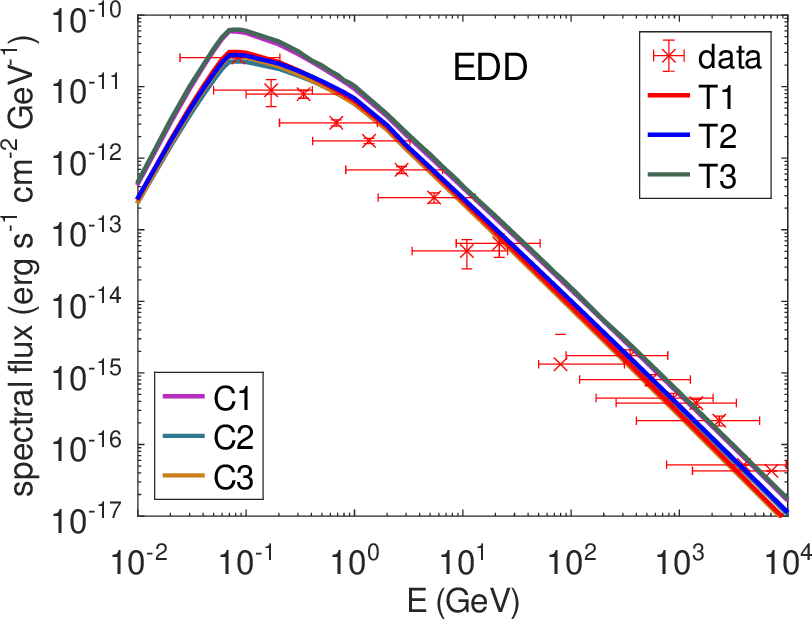,width=7.2cm,height=7.2cm,clip=}
\hskip -1mm
\epsfig{file=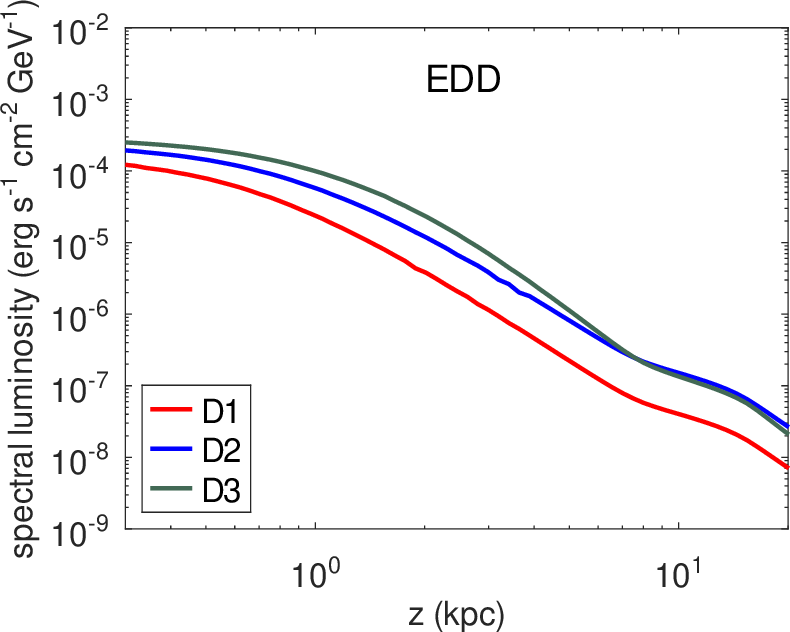,width=7.2cm,height=7.2cm,clip=}\\
\epsfig{file=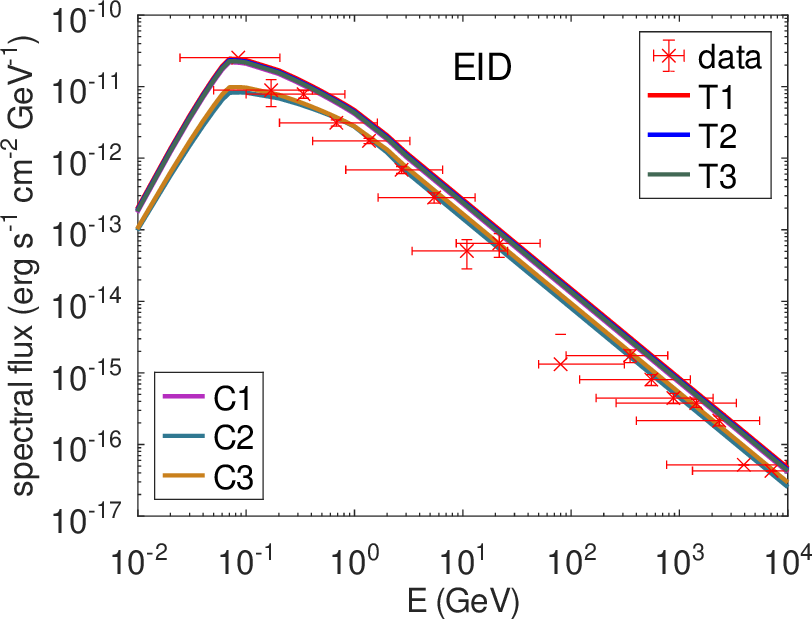,width=7.2cm,height=7.2cm,clip=}
\hskip -1mm
\centering
\epsfig{file=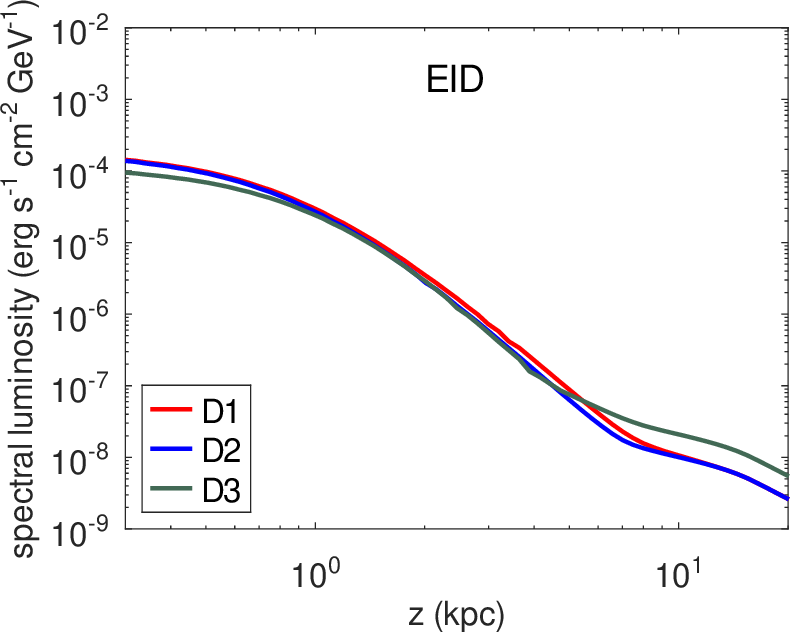,width=7.2cm,height=7.2cm,clip=}
\caption{Upper and lower left-hand panels: calculated 
spectral $\gamma$ flux of the CSB (cases C1--C3) and disc-halo including the CSB (cases T1--T3), for energy-dependent (top) and independent (bottom)
diffusion. Also shown are the
$\gamma$-fluxes measured by H.E.S.S. and \textit{Fermi}/LAT (Abdalla \ea 2018). Upper and
lower right-hand panels: los-integrated $\gamma$ -emissivity profiles at 100 MeV as a function of height above the disc plane, for energy-dependent (top) and independent (bottom) diffusion.}
\label{fig:gamma}
\end{figure}

The spectral $\gamma$-luminosity of the CSB and disc-halo in the 1 MeV--10 TeV
range are shown in Fig.~\ref{fig:gamma} for the EDD (upper) and EID (lower) models. We
emphasize that with the particle distributions normalized by the radio measurements,
and the selected (case-dependent) gas densities, 
the predicted $\gamma$-ray spectra are fully determined. The figures include the 
spectral flux measured by H.E.S.S. and \textit{Fermi}/LAT in the 60 MeV--5.75 TeV range
(Abdalla \ea 2018). Cases T1--T3 are the sum of the CSB case C2 and the respective
disc-halo emission. Los-integrated spectral $\gamma$ emissivities at different heights
above the disc plane are shown in the right-hand panels of the figure. As is clear from
the figures, the emission from the CSB constitutes a significant fraction of the total 
disc emission; specifically, the central region contributes 80--90 per cent of the
integrated 10 MeV--10 TeV in both EID and EDD models. Still, for viable range of values
of key disc parameters, $\gamma$-ray emission from the rest of the disc has to be
accounted for, particularly at low energies (below $\sim 10$ GeV).

As mentioned in the previous section, the AREPO MHD simulations of Werhahn \ea 
(2021) specifically included a galaxy whose basic properties were chosen to resemble 
those of NGC253. Model spectra of the hadronic \gr emission produced by a steady-state 
distribution of energetic protons were normalized to the \textit{Fermi}/LAT-H.E.S.S. observations.
As such, their spectra are roughly similar to our calculated spectra, which fit the (same) 
data reasonably well. In this (specific) regard the main difference between our 
phenomenological approach to their simulated-based study stems from the fact 
that we normalize the particle spectral distributions by comparison of their (combined) 
synchrotron emission with the good-quality radio spectra of Kapi\'{n}ska \ea (2017).

We note in passing that (as has already been noted in several previous studies) the leptonic contributions to $\gamma$-ray emission by (NT) bremsstrahlung and Compton scattering are small, particularly so at energies below 100 GeV (\eg, Rephaeli, Arieli, and Persic 2010, Werhahn \ea 2021).

Finally, whereas there is little variation in the EDD and EID spectra in the respective 
C1--C3 and T1--T3 cases, at high energies the EDD spectra are somewhat steeper than the corresponding EID spectra. Specifically, the slope of the high-energy part of the EDD spectra of both the CSB and total flux is $\sim 1.49-1.52$, as compared with $1.25$ for the EID models, which, as expected,  retraces the power index of the proton progenitors. Clearly, the faster EDD of high-energy protons out of the central, denser regions accounts for the steepening of the spectrum at the higher $\gamma$-energies. Note that in
their study of Galactic $\gamma$-emission from $\pi^0$-decay Aharonian \& Atoyan 
(1996) infer a scaling relation between the power-law index of the $\gamma$-ray spectral index and the indices of the proton injection spectrum and that of the diffusion coefficient, 
$\alpha'=\alpha+(3/2)\delta$, valid at high ($\gamma$-ray) energies. This relation 
would imply $\alpha'=1.7$ for $\alpha=1.25$ and $\delta=0.3$, higher by 40 per cent
than our deduced value. The
difference can be explained by the fact that particle sources in our treatment are distributed throughout the disc, whereas Aharonian \& Atoyan (1996) assumed a point-source model. Clearly, the spectrum is flatter when at any point in the disc the proton population includes freshly injected protons in addition to previously injected (thus, energy-degraded) protons.

\section{Discussion}

Studies of SBGs are important from a general point of view, due to the basic role of these galaxies in galactic evolution histories, and more specifically (and -- from observational point of view -- practically), in light of the enhanced stellar-driven phenomena that are radiatively manifested across the electromagnetic spectrum. These phenomena include the acceleration
of electrons and protons and their interactions with the magnetized interstellar medium and local radiation fields, resulting in high levels of radio, X-and-$\g$-ray emission. Moreover, given that SBGs are relatively abundant in the broad class of active galaxies, their high energy emission may account for a significant fraction of the observed diffuse X-ray (e.g., Persic and Rephaeli 2003) and-$\g$-ray (Roth \ea 2021) backgrounds.

The main objective
of our extended spectro-spatial modelling of the nearby NGC253 has been to
determine separately the distributions of energetic electrons and protons, and their 
radiative yields in the CSB, disc and halo regions. In the CSB region, all relevant stellar
(SN rate, SFR), gas, magnetic and radiation fields properties are clearly much more pronounced than in the rest of the disc, a fact that justifies modelling the CSB separately,
in spite of its relatively small volume. Indeed, radio emission from the central region comprises an appreciable fraction of the total galactic emission, as is clear from the 
measured spectra shown in figure 4 (and tables 4 \& 5) of Kapi\'{n}ska \ea (2017). 
These spectro-spatial radio measurements provide the main observational basis in our 
study.

The underlying assumption in our treatment has been that a low-luminosity AGN 
powered by accretion on to a MBH does not contribute appreciably to the
measured CSB emission, particularly so to the level of the measured radio flux which is 
used to determine the overall normalization factor. Given the various modelling and
inherent observational uncertainties stemming from the limited spatial resolution of the CSB 
radio emission, the practical consideration is whether the MBH contributes at a level which 
is more than (roughly) $20\%$ to the measured flux. As noted in the Introduction, the selection 
of NGC253 for this study is partly based on the expectation that in this SBG (unlike M82) the 
relative contribution of the central source is small both in radio (as can be deduced from the 
work of Kapi\'{n}ska \ea 2017) and also in \gr (Guti\'{e}rrez \ea 2020). Nevertheless, the main
impact of an appreciable contribution of a central source on our quantitative results would amount to somewhat reduced values of the overall normalization factors specified in the previous section.

As we have previously noted (in the Introduction), the diffusion-based modelling of energetic
particle propagation in IS space is meant to be an approximate description of what realistically 
is a more complex mode of transport that involves also advection, possibly as part of (a large 
scale) galactic wind. Clearly, inclusion of the latter propagation mode introduces additional 
free parameters; this is unwarranted given the current level of precision of the observational 
data base. By including spatial and energy dependence of the diffusion coefficient we can (at
least partly) gauge the impact of advection with flow speed that varies with (vertical)    
distance from the disc through a corresponding increase in the diffusion coefficient. For example, in the simulations of Thomas \ea (2023), the wind speed reaches high values
($> 300$ km s$^{-1}$) only at large distances from the disc, whose possible impact is partly accounted for by the high values of the diffusion coefficient in our models.

NT X-ray emission is mainly by Compton scattering of energetic electrons off the intense 
optical and FIR radiation fields in the disc. Our estimates of the predicted emission in the
7--20 keV band are below the bounds from NuStar measurements set by Wik \ea (2014;
table 3) on diffuse (point-source subtracted) emission in the central disc region.
However, the range of models considered here, the predicted full-disc emission is
considerably higher than that of the CSB.

For the range of the EID models considered in our work, the spectral (energy) index of the
\gr emission at energies higher than 100 MeV (i.e., beyond the $\pi^{0}$-decay peak) is $1.0--1.25$ for emission from the CSB and also for emission from the rest of the disc.
These values are roughly in the range deduced by Abdalla \ea (2018) from \textit{Fermi}/HESS
measurements, $\sim 1.4 \pm 0.3$. The predicted spatial profile of the disc \gr emission is
quite steep; at $\sim 1$ kpc the flux drops by a factor of 2--3 (for all three cases) with
respect to its value at the boundary of the CSB.  As we have already noted (in the 
previous section), most of the disc emission in the (combined \textit{Fermi}/LAT-H.E.S.S.) 10 MeV--10 TeV flux, a fraction of $\sim 0.8$--$0.9$, originates in the CSB. However, CSB
emission clearly does not dominate the total galactic emission, neither in radio nor in 
hard X-ray. Therefore, modelling the galactic emission as originating largely from the
small CSB region (as assumed in some previous works) seems quite unrealistic. Also, 
for the range of realistic values of the key parameters assumed here, NT emission 
from the galactic halo is practically negligible. While this is expected, our predicted 
\gr spatial profile can be useful for estimating the size of the inner halo region that has 
to be observed when detection of the full galactic emission is sought. 

The result that a significant part of the \gr emission in NGC253 originates outside the 
CSB region has theoretical and observational implications. It clearly indicates that 
fitting the measured spectrum by modelling only the CSB can lead to significant bias in
the basic derived quantities, especially those characterizing the particle spectra and gas 
densities in this region. Nuclear SB regions of nearby SBGs have been studied in several previous works (cited in the Introduction) and more recently by Peretti \ea (2019), whose 
main objective was to determine the conditions for energetic particle confinement in the nuclear SB regions of M82, NGC253 and Arp220. To do so they considered particle propagation to be dominated either by diffusion or advection, and compared their 
predicted spectral models with X-and-\gr measurements, assuming that these originate 
mostly in the small nuclear SB region (with assumed radius of 200 pc). In contrast, our 
work here has been motivated by the realization, based on previous work (e.g., Persic \ea 
2008; Rephaeli \ea 2010) that emission from the full galactic disc has to be modelled and accounted for when comparison is made with multispectral measurements. The results presented in the previous section strengthen this conclusion. Therefore, comprehensive modelling of particle and radiative
yields in the disc has to be based on a spectro-spatial analysis, as has been done here.

The ratio of the particle total (mostly radiative) energy flux resulting from interactions in 
IS space to the particle energy flux emerging from their acceleration sites is referred to as 
the calorimetric efficiency. Estimates of this fraction are typically high ($\sim 90$ per cent) for
primary electrons, but can extend down to $\sim 30$ per cent for protons, as reported by Krumholz \ea (2020)
for NGC253 (and M82). Obviously, additional losses in the halo result in essentially negligibly small residual primary electron component in the outer halo. Of interest is mainly the total calorimetric efficiency of protons and their residual energy content in the outer halo. For
the models explored here, our estimates indicate that the total calorimetric efficiencies 
of protons in the 10 MeV--10 TeV band are in the range
$\sim 30-50$ per cent (EID, cases 2 and 1) and $\sim 50-80$ per cent (EDD, cases 3 and 1).
These estimates are based on our assumed parameter values in the three regions, 
including values of the magnetic field, diffusion coefficient, and gas mass in the halo. 
The substantial uncertainty in the values of the latter two quantities in the halo can significantly affect estimates of energetic proton losses in this vast region, thereby 
impacting the calorimetric efficiency and the contribution of SFGs to the proton energy 
flux in IG space. The highest estimated calorimetric efficiencies (case 1, in both EID and 
EDD models) are obtained for the lower value of the central diffusion coefficient
(in both the CSB and disk), as well as with lower $\beta_D$, for which the 
growth of the diffusion profile is milder. The resulting slower proton diffusion (and consequently the more frequent interactions with ambient protons) in these denser 
regions enhances (the pion-decay) $\gamma$-ray yield, and thus also the respective calorimetric efficiencies. This dependence of the calorimetric efficiency on the diffusion coefficient has already been deduced in previous works (e.g., Werhahn \ea 2021).

Reliable estimates of the calorimetric efficiency of protons in SFGs are needed in the 
study of NT phenomena in galaxy clusters. In the central regions of rich clusters the gas density and magnetic fields are significantly enhanced with respect to their levels in IG 
space outside clusters, with typical values of O(10$^{-3}$) cm$^{-3}$ and O(1) 
$\mu$G, higher than corresponding values in the outer halo of a SFG that is not a cluster member. The higher concentration of SFGs in a cluster and stronger coupling of protons 
to the magnetized intracluster gas result in the creation of secondary electrons and the 
emission of significant radio and NT X-\&-$\gamma$ emission in the central cluster 
region (Rephaeli and Sadeh 2016).

\section{Acknowledgments}
We are grateful to the referee for a constructive report and suggestions that have widened the scope of our work. This research has been supported by a grant from the Joan and Irwin Jacobs donor-advised fund at the JCF (San Diego, CA).

\section*{Data Availability}
The data underlying our work are available in the manuscript.

\section{REFERENCES}
\def\ref{\par\noindent\hangindent 20pt}

\ref Abdalla H. \ea, 2018, \Aa, 617, A73
\ref Abdo, A. A. \ea, 2010a, \Aa, 523, L2
\ref Abdo, A. A. \ea, 2010b, \apj, 709, L152
\ref Abdo A.A. et al., 2010, \apj, 719, 1433
\ref Acciari V.A., et al. (VERITAS Collab.) 2009, Nature, 462, 770 
\ref Acero F. et al. (HESS Collab.) 2009 Science, 326, 1080 
\ref Atoyan A.M., Aharonian F.A., \& V\"{o}lk H.J., 1995, \prd, 52, 3265
\ref Blumenthal G.R. \& Gould R.J., 1970, \rmph, 42, 237
\ref de Cea del Pozo E., Torres D.F., Rodriguez Marrero A.Y., 2009, ApJ, 698, 1054
\ref Chan T.K., Kere\v{s} D., Hopkins P.F., Quataert E., Su K.-Y., Hayward C.C.,
Faucher-Gigu\`{e}re C.-A., 2019, \mn, 488, 3716
\ref Domingo-Santamar\'{i}a E., Torres D.F., 2005, A\&A, 444, 403
\ref Evoli C., Blasi P., Amato E., Aloisio R., 2020, \prl, 125, 051101
\ref Goldshmidt O., Rephaeli Y., 1995, \apj, 444, 113
\ref Guti\'{e}rrez E.M., Romero G.E., Vieyro F.L., 2020, \mn, 494, 2109
\ref Heesen V., Beck R., Krause M., Dettmar R.-J., 2009, \Aa, 494, 563
\ref Heesen V. et al., 2018, \mn, 476, 158
\ref Hopkins P.F et al., 2020, \mn, 492, 3465
\ref Kapi\'{n}ska A.D. \ea, 2017, \apj, 838, 68
\ref Krumholz M.R., Crocker R.M., Xu S., Lazarian A., Rosevear M.T., Bedwell-Wilson J., 2020,
\mn, 493, 2817
\ref Lacki B.C., Thmopson T.A., Quataert E., Loeb A., Waxman E., 2011, \apj, 734, 107
\ref Lucero D.M., Carignan C., Elson E.C., Randriamampandry T.H., Jarrett T.H.,
Oosterloo T.A., Heald G.H., 2015, \mn, 450, 3935
\ref Mannheim K. \& Schlickeiser R., 1994, \Aa, 286, 983
\ref Moskalenko I.V., Jones F. C., Mashnik S. G., Ptuskin V. S., Strong A. W.,
2003, in Kajita T., Asaoka Y., Kawachi A., Matsubara Y., Sasaki M.,
eds, International Union of Pure and Applied Physics (IUPAP), 28th Int.
Cosmic Ray Conf., GALPROP: New Developments in CR Propagation
CodeShow Affiliations. p. 1925
\ref Paglione T.A.D, Marscher A.P., Jackson J.M., Bertsch D.L., 1996, \apj, 460, 295
\ref Peretti E., Blasi P., Aharonian F., Morlino G., 2019, \mn, 487, 168
\ref Persic M. Rephaeli Y., 2003, A\&A, 399, 9
\ref Persic M. Rephaeli Y., Arieli Y., 2008, A\&A, 486, 143
\ref Rephaeli Y., Arieli Y., Persic M., 2010, \mn, 401, 473
\ref Rephaeli Y., Sadeh S., 2016, \prd, 93, 101301
\ref Rephaeli Y., Sadeh S., 2019, \mn, 486, 2496
\ref Roth M. A., Krumholz M. R., Crocker R. M., Celli S., 2021, Nature, 597, 341
\ref Ruszkowski M., Yang K.H.-Y., Reynolds C.S., 2017, \apj, 844, 13
\ref Strong A.W., Porter T.A., Digel S.W., J\'{o}hannesson G., Martin P., Moskalenko I.V., Murphy E.J., Orlando E., 2010, \apjl, 722, L58
\ref Thomas T., Pfrommer C., Pakmor R., 2023, \mn, 521, 3023
\ref Torres D.F., 2004, ApJ, 617, 966
\ref V\"{o}lk, H.J., Aharonian, F.A., Breitschwerdt, D., 1996, Sp. Sci. Rev, 75, 279
\ref Werhahn M., Pfrommer C.,  Girichidis P., Georg W., 2021, MNRAS, 505, 3295
\ref Werhahn M., Girichidis P., Pfrommer C.,  Wittingham J., 2023, MNRAS, 525, 4437
\ref Wiegert T. et al., 2015, Astron. J. 150, 81
\ref Wik D.R. \ea, 2014, \apj, 797, 79
\ref Zweibel E.G., 2013, \php, 20, 055501

\end{document}